\documentclass[conference]{IEEEtran}
\IEEEoverridecommandlockouts
\usepackage{cite}
\usepackage{verbatim}
\usepackage{pgfplots}
\pgfplotsset{compat=newest}
\usetikzlibrary{plotmarks}
\usetikzlibrary{arrows.meta}
\usepgfplotslibrary{patchplots}
\usepackage{grffile}
\usepackage{amsmath,amssymb,amsfonts}
\usepackage{hyperref}
\usepackage{cleveref}
\usepackage{tikz}
\usepackage{algorithmic}
\usepackage{graphicx}
\usepackage{subfigure}
\usepackage{epstopdf}
\usepackage[utf8]{inputenc}

\def\BibTeX{{\rm B\kern-.05em{\sc i\kern-.025em b}\kern-.08em
    T\kern-.1667em\lower.7ex\hbox{E}\kern-.125emX}}

\newcommand\copyrighttext{%
  \footnotesize \textcopyright 2022 IEEE. Personal use of this material is permitted. Permission from IEEE must be obtained for all other uses, in any current or future media, including reprinting/republishing this material for advertising or promotional purposes, creating new collective works, for resale or redistribution to servers or lists, or reuse of any copyrighted component of this work in other works.}
\newcommand\copyrightnotice{%
\begin{tikzpicture}[remember picture,overlay]
\node[anchor=south,yshift=10pt] at (current page.south) {\fbox{\parbox{\dimexpr\textwidth-\fboxsep-\fboxrule\relax}{\copyrighttext}}};
\end{tikzpicture}%
}

\begin{document}

\title{On the Modeling and Analysis of Fast Conditional Handover for 5G-Advanced \vspace{-0.5\baselineskip} }

\author{\IEEEauthorblockN{Subhyal Bin Iqbal$^{*\dagger}$, Ahmad Awada$^{*}$, Umur Karabulut$^{*}$, Ingo Viering$^{*}$, Philipp Schulz$^{\dagger}$ and Gerhard P. Fettweis$^{\dagger}$}
\IEEEauthorblockA{$^{*}$Standardization and Research Lab, Nokia, Munich, Germany} \
{$^{\dagger}$Vodafone Chair for Mobile Communications Systems, Technische Universität Dresden, Germany \vspace{-1.2\baselineskip}}\\
}

\maketitle

%
\copyrightnotice

\begin{abstract}
Conditional handover (CHO) is a state-of-the-art 3GPP handover mechanism used in 5G networks. Although it improves mobility robustness by reducing mobility failures, the decoupling of the handover preparation and execution phases in CHO significantly increases the signaling overhead. For 5G-Advanced networks, fast CHO (FCHO) is a recent 3GPP proposal that offers a practical solution whereby the user equipment (UE) can reuse earlier target cell preparations after each handover to autonomously execute subsequent handovers. This saves the signaling overhead associated with the reconfiguration and repreparation of target cells after each handover. In this paper, a comprehensive study on the mobility performance of FCHO with respect to mobility failures and signaling overhead in frequency range 2 (FR2) is carried out. In particular, the performance of FCHO is compared with CHO for two different multi-panel UE (MPUE) schemes. Results show that FCHO substantially reduces  the signaling overhead of CHO, while at the same time it also reduces mobility failures due to faster triggering of the handover that is achieved by saving the preparation delay.
\end{abstract}

\vspace{-0.58\baselineskip}
\section{Introduction} \label{Sec1}
Although frequency range 2 (FR2) \cite{b1} offers a practical solution to the problem of contiguous bandwidth that is needed for 5G mobile networks to fulfill the steep increase in user demand, it introduces further challenges to radio propagation such as higher free-space path loss and penetration loss \cite{b2}. This leads to rapid signal degradation in mobile environments where there are many static and moving obstacles. It also particularly pronounces the problem of inter-cell interference \cite{b3}, which besides being high in cell boundary regions is now also subject to rapid increase. Hence, a high number of mobility failures are experienced.

Conditional handover (CHO) has been introduced in \cite{b4} as an alternate to baseline handover to improve mobility robustness. In CHO, handover preparation and execution is decoupled by introducing a conditional procedure. Hereby, the handover towards the prepared target cell is prepared early by the serving cell but random access is performed later by the user equipment (UE), when the link quality is sufficient. However, even if the mobility parameters chosen from a search space yield optimal mobility performance, CHO still brings about the problem of a significant signaling overhead \cite{b3}. In FR2, particularly there is a very high signaling overhead due to a high number of handovers due to dense cell deployment. 

For 5G-Advanced networks \cite{b4.5}, an alternate handover mechanism to address this issue is under discussion \cite{b5, b6}. Fast CHO (FCHO) permits the UE to retain its prepared target cells after a handover, and later reuse them to autonomously execute handovers. This significantly reduces the signaling overhead by saving the preparation of multiple target cells. At the same time mobility failures are also reduced because handovers are now prepared relatively faster and can be executed immediately when a better cell becomes available. 

On the UE architecture side, another solution that has been proposed is multi-panel UE (MPUE), i.e., a UE equipped with more than one spatially distinct antenna panels \cite{b7, b8}. It offers a higher directional gain on each panel as compared to an isotropic UE that has a single antenna panel with an isotropic radiation pattern. Furthermore, if the MPUE can be made to communicate with one or more panels oriented towards the serving cell then the inter-cell interference from the neighboring cells can be significantly suppressed.

In this paper, the mobility performance of FCHO in terms of mobility key performance indicators (KPIs) and signaling overhead is investigated. The mobility performance is elaborated for two different MPUE signal measurement schemes that are addressed in 3GPP standardization \cite{b9}. It is important to understand the mobility performance of FCHO for advanced UE architectures since both are an essential part of 5G-Advanced \cite{b4.5}. To the best of the authors' knowledge, the mobility performance of FCHO has not been investigated in literature before. The rest of the paper is organized as follows. In \Cref{Sec2}, CHO preparation, release, and replace events are explained along with the signaling which is required for each CHO event. In \Cref{Sec3}, FCHO mechanism is introduced and the main benefits against conventional CHO are highlighted. In \Cref{Sec4}, the simulation scenario and the two different MPUE signal measurement schemes are explained. In \Cref{Sec5}, the mobility KPIs and CHO signaling events which reflect the signaling overhead are explained. Then in \Cref{Sec5} the performance of FCHO and CHO are analyzed for the MPUE and isotropic UE architectures for two realistic mobility scenarios. Finally, in \Cref{Sec6} the paper is concluded along with an outlook for future enhancements. 

\section{Conditional Handover} \label{Sec2}

In this section, the CHO mechanism and the three distinct signaling events that trigger UE measurement reports in CHO are discussed in detail. These are the CHO preparation, release, and replace events.

\subsection{CHO Preparation Event} \label{Subsec2.1}
Each UE in the network is configured by the serving cell to measure the raw reference signal received power (RSRP)  $P_{c,b}^\textrm{RSRP}(n)$ (in dBm) at a discrete time instant $n$ from each beam $b \in B$ of cell $c \in C$, using the synchronization signal block (SSB) bursts transmitted by each cell. One or more such cells are controlled by a gNodeB (gNB) and it is assumed that the serving cell $c_0$ and a prepared target cell $c^{\prime}$ are controlled by different serving and target gNBs, namely $\textrm{gNB}_0$ and $\text{gNB}^\prime$. The raw RSRPs undergo layer 1 (L1) and L3 filtering at the UE to mitigate the effect of channel impairments. The L3 filtered RSRP is defined as $P_{c,b}^\textrm{L3}(m)$, where $m=n\omega$ and $\omega \in \mathbb{N}$ is the SSB periodicity. The L3 cell quality measurements derived after beam consolidation are then $P_{c}^\textrm{L3}(m)$ by the UE for handover decisions \cite{b7}.

The CHO  \textit{preparation} event is needed so that the UE can prepare target cell $c^{\prime}$ for handover. It follows the CHO preparation condition that is monitored by the UE connected to serving cell $c_0$ and defined as

\vspace{-\baselineskip}
\begin{equation}
\label{Eq1}
     P_{c_0}^\textrm{L3}(m)  < P_{c^{\prime}}^\textrm{L3}(m) + o^\mathrm{prep}_{c_0,c^{\prime}} \ \text{for} \  m_\textrm{prep} - T_\mathrm{prep} < m < m_\textrm{prep},
\end{equation}

where $o^\mathrm{prep}_{c_0,c^{\prime}}$ is defined as the CHO preparation offset between cell $c_0$ and $c^{\prime}$. The UE sends a measurement report to serving cell $c_0$ at time $m=m_\mathrm{prep}$ if the preparation condition is fulfilled for the preparation condition monitoring time $T_\mathrm{prep}$. 

As mentioned in \Cref{Sec1}, CHO handover execution is decoupled from CHO preparation whereby the handover is prepared early but the actual handover execution is performed only when the radio link is sufficient. The UE monitors the CHO execution condition, defined as

\vspace{-\baselineskip}
\begin{equation}
\label{Eq2}
     P_{c_0}^\textrm{L3}(m) + o^\mathrm{exec}_{c_0,c^{\prime}}  < P_{c^{\prime}}^\textrm{L3}(m) \ \text{for} \ m_\textrm{exec} - T_\mathrm{exec} < m < m_\textrm{exec},
\end{equation}
where $o^\mathrm{exec}_{c_0, c^{\prime}}$ is defined as the CHO execution offset between cell $c_0$ and $c^{\prime}$. The UE executes a handover towards the prepared target cell $c^{\prime}$ if the execution condition at $m=m_\mathrm{exec}$ is fulfilled for the execution condition monitoring time $T_\mathrm{exec}$.

\begin{figure} [!t]
\centering
\includegraphics[width = 0.97\columnwidth, keepaspectratio]{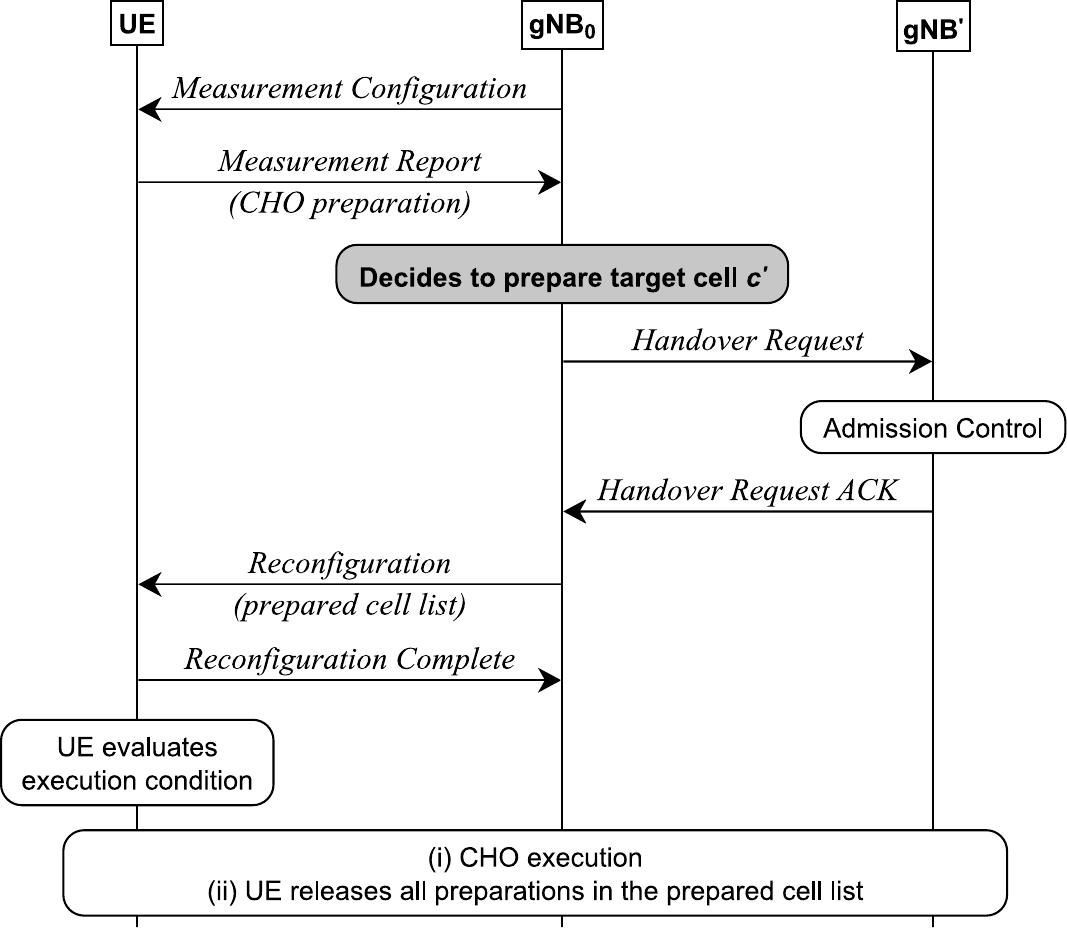}
\vspace{-0.3\baselineskip}
\caption{CHO preparation event signaling diagram followed by CHO execution.}
\label{Fig1} \vspace{-0.9\baselineskip}
\end{figure}

Fig. \ref{Fig1} shows the signaling diagram between the UE and the network for the CHO preparation event followed by CHO execution. The UE constantly monitors the preparation condition, configured by the \textit{Measurement Configuration} message. Once the preparation condition in (\ref{Eq1}) is satisfied for the prepared target cell $c^{\prime}$, CHO preparation is triggered. Thereafter,  the measurement report is sent by the UE to $\textrm{gNB}_0$ that controls the serving cell $c_0$. $\textrm{gNB}_0$ sends a \textit{Handover Request} message to $\textrm{gNB}^\prime$ that controls the prepared target cell $c^{\prime}$. $\textrm{gNB}^\prime$ performs admission control and acknowledges (\textit{ACK}) the handover request. Thereafter, $\textrm{gNB}_0$ sends a \textit{Reconfiguration} message to the UE and it adds cell $c^{\prime}$ to its prepared cell list. The UE acknowledges this with a \textit{Reconfiguration Complete} message. If there are multiple target cells that fulfill the preparation condition in (\ref{Eq1}), then multiple target cells are prepared. The UE continues its connectivity to $\textrm{gNB}_0$ but herein it constantly monitors the CHO execution condition in (\ref{Eq2}) for all the prepared target cells. When the execution condition is fulfilled for any cell, the UE detaches from $\textrm{gNB}_0$ and initiates a handover using random access towards $\textrm{gNB}^\prime$. Upon completion of a successful handover, $\textrm{gNB}^\prime$ sends an ACK to  $\textrm{gNB}_0$. Thereafter, the UE releases all prepared target cell configurations.

\subsection{CHO Release Event} \label{Subsec2.2}
In case the RSRP of any prepared target cell $c^{\prime}$ degrades after preparation, the resources that are allocated for handover by cell $c^{\prime}$ for that particular UE should be released so that they can be accessed by other UEs in the network. This ensures resource efficiency. But before this occurs, the UE releases the measurement configuration of cell $c^{\prime}$ so that the execution condition in (\ref{Eq2}) is no longer monitored. This is defined as the CHO \textit{release} event and is triggered by the CHO release condition. The CHO release condition, along with the CHO execution condition is given to the UE by the serving cell $c_0$ when cell $c^{\prime}$ is prepared. It is defined as

\vspace{-\baselineskip}
\begin{equation}
\label{Eq3}
     P_{c^{\prime}}^\textrm{L3}(m) + o^\mathrm{rel}_{c_0,c^{\prime}}   < P_{c_0}^\textrm{L3}(m) \ \text{for} \ m_\textrm{rel} - T_\mathrm{rel} < m < m_\textrm{rel},
\end{equation}

where $o^\mathrm{rel}_{c_0,c^{\prime}}$ is defined as the CHO release offset between cell $c_0$ and $c^{\prime}$. The UE sends a measurement to the serving cell $c_0$ at  $m=m_\mathrm{rel}$ if the release condition is fulfilled for the release condition monitoring time $T_\mathrm{rel}$. 

\begin{figure} [!b]
\vspace{-1.2\baselineskip}
\centering
\includegraphics[width = 0.81\columnwidth, height=0.49\columnwidth, keepaspectratio]{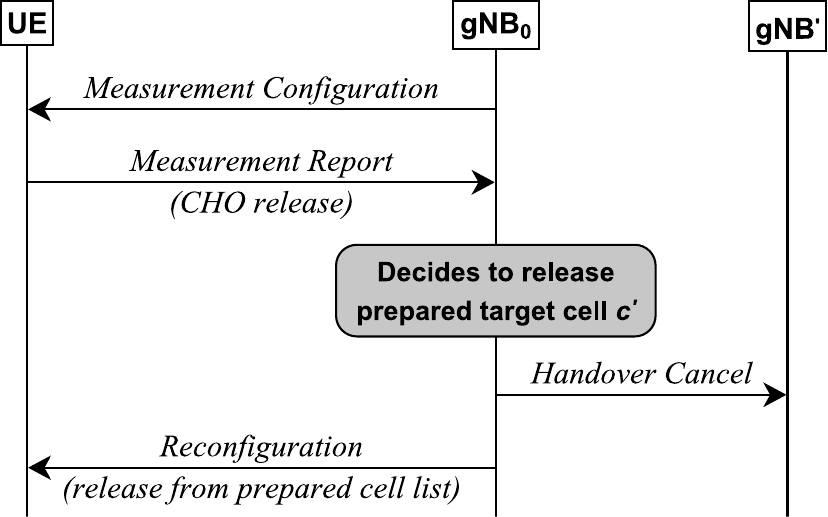}
\vspace{-0.3\baselineskip}
\caption{CHO release event signaling diagram.}
\label{Fig2} 
\end{figure}

Fig. \ref{Fig2} shows the signaling diagram between the UE and the network for the CHO release event. The UE constantly monitors the release condition which is configured by the \textit{Measurement Configuration} message. Once the release condition in (\ref{Eq3}) is satisfied for cell $c^{\prime}$, CHO release is triggered and the measurement report is sent by the UE to $\textrm{gNB}_0$ to initiate the CHO release procedure. $\textrm{gNB}_0$ communicates the \textit{Handover Cancel} message to $\text{gNB}^\prime$ and then confirms the preparation removal by $\textrm{gNB}^\prime$ via a \textit{Reconfiguration} message to the UE, which then releases the configuration of cell $c^{\prime}$. Such that frequent preparation and release for the same target cell do not occur, a hysteresis offset $o^\mathrm{hys}_{c_0, c^{\prime}}$ between $o^\mathrm{prep}_{c_0,c^{\prime}}$ and $o^\mathrm{rel}_{c_0,c^{\prime}}$ is defined, i.e., $o^\mathrm{rel}_{c_0,c^{\prime}} = o^\mathrm{prep}_{c_0,c^{\prime}} + o^\mathrm{hys}_{c_0,c^{\prime}}$ for any $o^\mathrm{hys}_{c_0,c^{\prime}} > 0$ \cite{b3}.

\subsection{CHO Replace Event} \label{Subsec2.3}
The list of prepared target cells has a limited number of entries in order to reduce the signaling overhead from the CHO preparation events and also to limit the prepared target cell resource reservations \cite{b3}.  To minimize mobility failures, it is essential that the prepared cell list is kept up-to-date even when all the entries are full. Therefore, the weakest prepared cell in the list $c_\textrm{W}$ can be replaced by another stronger neighboring cell $c_\textrm{S}$ through the CHO replace condition, defined as

\begin{figure} [!b] \vspace{-1.3\baselineskip}
\centering
\includegraphics[width = 0.88\columnwidth, height=0.51\columnwidth, keepaspectratio]{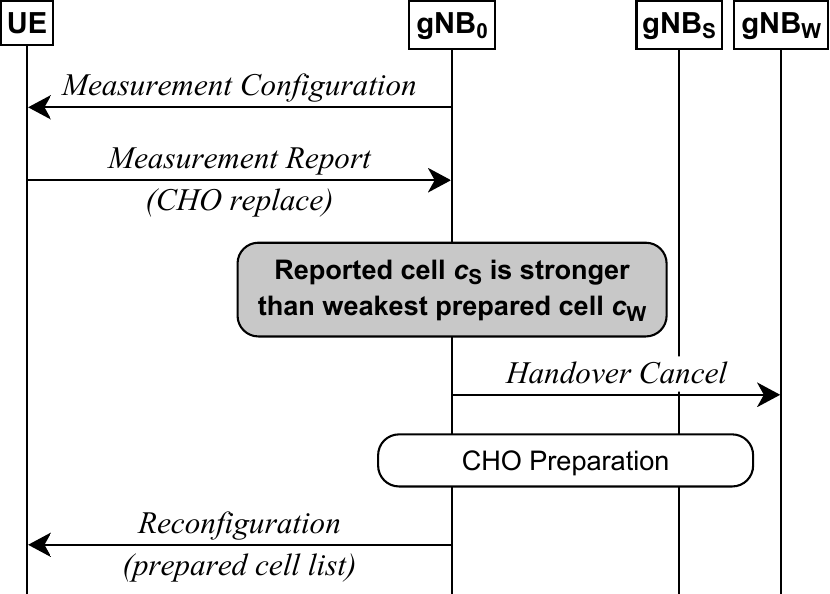}
\vspace{-0.4\baselineskip}
\caption{CHO replace event signaling diagram.}
\label{Fig3} 
\end{figure}

\vspace{-\baselineskip}
\begin{equation}
\label{Eq4}
     P_{c_\textrm{S}}^\textrm{L3}(m) > P_{c_\textrm{W}}^\textrm{L3}(m) + o^\mathrm{rep}_{c_\textrm{W,S}} \ \text{for} \ m_\textrm{rep} - T_\mathrm{rep} < m < m_\textrm{rep},
\end{equation}
where $o^\mathrm{rep}_{c_\textrm{W,S}}$ is defined as the CHO replace offset between cel $c_\textrm{W}$ and $c_\textrm{S}$. The UE sends a measurement report to the serving cell $c_0$ at time $m=m_\mathrm{rep}$ if the replace condition is fulfilled for the replace condition monitoring time $T_\mathrm{rep}$. The CHO replace condition, like the CHO release condition, is given to the UE by the serving cell $c_0$ when cell $c_\textrm{W}$ is prepared. Fig. \ref{Fig3} shows the signaling diagram between the UE and the network for the CHO replace event. Once the replace condition in (\ref{Eq4}) is satisfied for cell $c_\textrm{W}$, CHO replace is triggered and the measurement report is sent by the UE to $\textrm{gNB}_0$. Thereafter, $\textrm{gNB}_0$ sends a \textit{Handover Cancel} message to $\text{gNB}_\text{W}$ that controls cell $c_\textrm{W}$. Cell $c_\textrm{S}$ is then prepared based on the signaling outlined in Fig. \ref{Fig1}, and upon completion $\textrm{gNB}_0$ sends a \textit{Reconfiguration} message to the UE, which then replaces cell $c_\textrm{W}$ by $c_\textrm{S}$.

\section{Fast Conditional Handover} \label{Sec3}

In this section, the FCHO mechanism is explained along with its advantages in terms of mobility robustness and signaling overhead reduction. 

In \cite{b5, b6} it has been agreed that a handover mechanism, namely FCHO, can be defined where it “might be possible to keep CHO candidates after the handover” and reuse target cell preparations instead of releasing them after a handover. Two advantages are stated as such. Firstly, FCHO reduces mobility failures in certain situations, e.g., cell boundaries where high and rapidly increasing inter-cell interference may impact the ability of the network to successfully receive the measurement report or to provide a handover command timely to the UE. Consequently mobility failures can occur. The reuse of target cell preparations means that the cells are prepared relatively early, compared to conventional CHO, and hence a handover can be executed immediately. Secondly, FCHO reduces the signaling overhead caused by CHO signaling events because multiple target cell need not be prepared after every handover.

\begin{figure} [!b] \vspace{-1\baselineskip}
\centering
\includegraphics[width = 0.99\columnwidth, height=1.14\columnwidth]{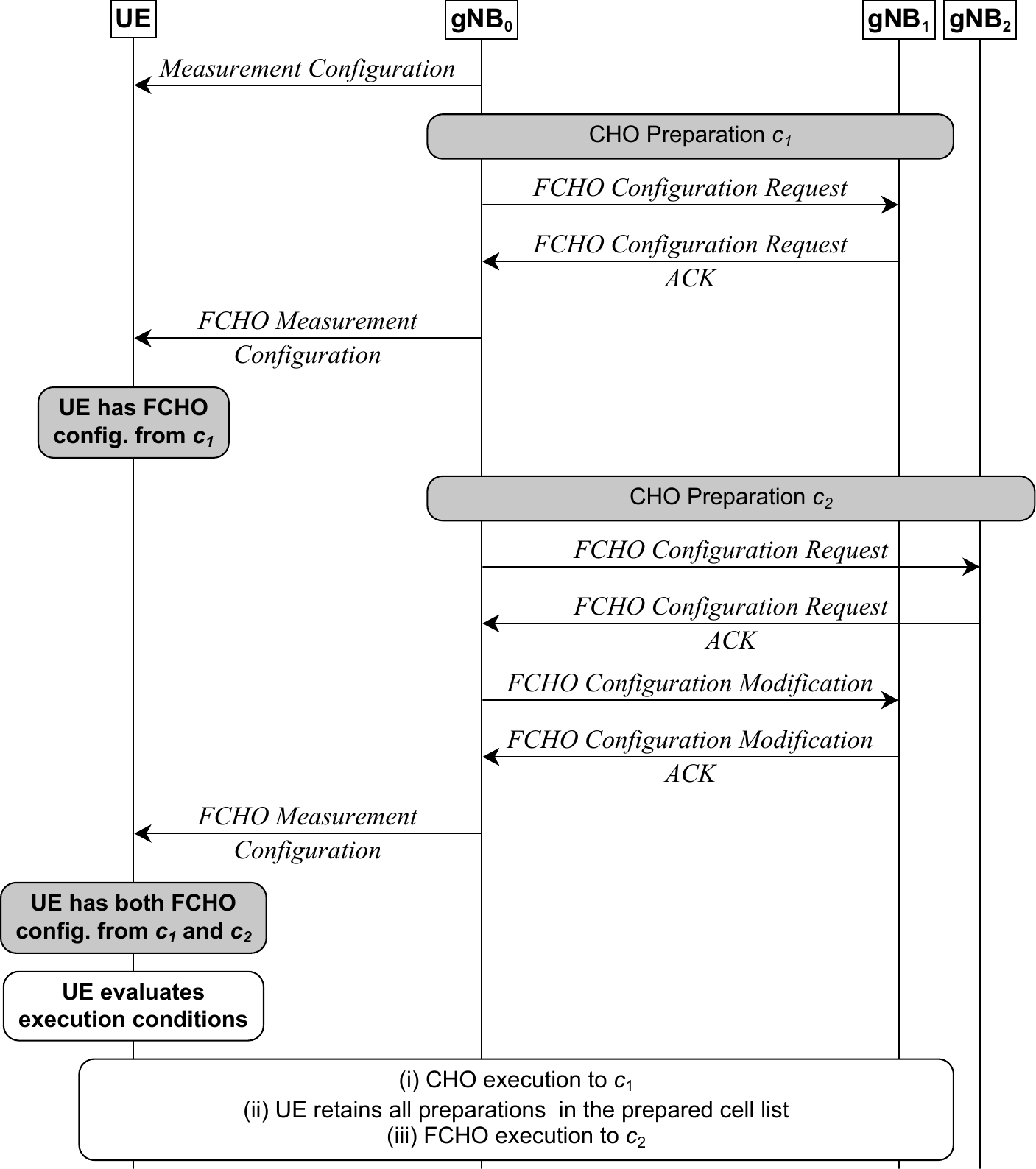} 
\vspace{-1.5\baselineskip}
\caption{FCHO signaling diagram.}
\label{Fig4} 
\end{figure}

 Fig. \ref{Fig4} shows the signaling diagram between the UE and the network for the FCHO mechanism, where the maximum number of prepared cells $n_c^{\mathrm{max}}$ is assumed to be two. It is seen that initially the UE is connected to the serving cell $c_0$, controlled by $\textrm{gNB}_0$. When the preparation condition in (\ref{Eq1}) is fulfilled for cell $c_1$, the UE sends a measurement report to cell $c_0$. The cell preparation for cell $c_1$ follows the signaling outlined earlier in Fig. \ref{Fig1}. Once $c_1$ is prepared, $\textrm{gNB}_0$ initiates an \textit{FCHO Configuration Request} message to $\textrm{gNB}_1$, requesting the configuration needed to execute an FCHO from cell $c_1$ back to cell $c_0$, i.e., the CHO execution offset $o^\mathrm{exec}_{c_1,c_0}$. $\textrm{gNB}_1$ acknowledges with an \textit{FCHO Configuration Request ACK} message which includes the respective configurations. $\textrm{gNB}_0$ then forwards the respective configuration to the UE as an \textit{FCHO Measurement Configuration} message. From now on, the UE can execute a CHO from cell $c_0$ to $c_1$ if the CHO execution condition in \ref{Eq2} is fulfilled for a handover to $c_1$, and a subsequent FCHO back to cell $c_0$ if the execution condition if fulfilled later on for cell $c_0$.

The serving cell $c_0$ then prepares cell $c_2$ when the preparation condition in (\ref{Eq1}) is fulfilled for it. Once $c_2$ is prepared, $\textrm{gNB}_0$ initiates another \textit{FCHO Configuration Request} message to $\textrm{gNB}_2$, requesting the configuration needed to execute an FCHO from cell $c_2$ back to cell  $c_0$, as well as from cell $c_2$ to the other prepared target cell $c_1$. This is acknowledged and the requested configuration is sent to $\textrm{gNB}_0$. $\textrm{gNB}_0$ then initiates an \textit{FCHO Configuration Modification} message to the other prepared target cell $c_1$ via $\textrm{gNB}_1$, this time requesting the configuration needed to execute an FCHO from cell $c_1$ to $c_2$, which is acknowledged upon receipt. $\textrm{gNB}_0$ then sends a second \textit{FCHO Measurement Configuration} message to the UE, with the respective configurations received from $\textrm{gNB}_1$ and $\textrm{gNB}_2$. At this point, the UE can execute a CHO from cell $c_0$ to either of the two prepared target cells and can follow this up by an FCHO either back to cell $c_0$ itself or to the other prepared target cell.

The UE constantly monitors the CHO execution given in (\ref{Eq2})  for both $c_1$ and $c_2$. Fig. \ref{Fig4} shows the case where $c_1$ satisfies the execution condition first, upon which the UE detaches from  $\textrm{gNB}_0$ and initiates a handover using random access towards $\textrm{gNB}_1$. Upon completion of a successful CHO, $\textrm{gNB}_1$ sends an acknowledgement to $\textrm{gNB}_0$. However, as opposed to conventional CHO, the UE now swaps cell $c_1$ with cell $c_0$ (from which it has just detached) and retains the other preparation of cell $c_2$  in the prepared cell list. This means that the UE can now immediately perform FCHO execution towards either cell $c_2$  or $c_0$. The CHO execution condition for cell $c_2$ is shown to be fulfilled first in Fig. \ref{Fig4} and the UE immediately executes an FCHO towards it without any additional signaling or preparation delay.

It is useful to retain the previous serving cell configuration after a handover because it could be the next target cell. Hence, a subsequent preparation can be potentially saved. The main benefit of FCHO is on cell boundaries where the RSRPs of neighboring cells are relatively close to each other ($<$ 5 dB). Hence, a relatively faster handover to one of these neighboring prepared cells can be executed. In conventional CHO it would mean first preparing one or more of such neighboring cells as a target cell, leading to a late preparation that could result in a mobility failure. Later, the UE can autonomously perform a handover to any prepared targetcell that satisfies the CHO execution condition in (\ref{Eq2}). 

In FCHO, CHO preparation events are reduced compared to conventional CHO because the prepared cell list is not reset after a successful handover. Since $ n_c^{\mathrm{max}}$ can be up to eight cells as per 3GPP \cite{b1}, this would mean that eight CHO preparation events can be saved after each successful handover. Less CHO preparations also mean less CHO removals. On the other hand, some prepared cells may become outdated and need to be removed. Similarly, as compared to conventional CHO there may be more CHO replace events because some prepared target cells become weak over time. However, it is known from \cite{b3} that CHO preparation events dominate CHO release and CHO replace events and are responsible for most of the signaling overhead. Therefore, the overall signaling overhead in FCHO will be much less than in conventional CHO.

\section{Simulation Scenario and Parameters} \label{Sec4}

In this section, the simulation scenario is explained along with the simulation parameters that are listed in \Cref{Table1}.

\begin{table}[!b]
\vspace{-0.8\baselineskip}
\renewcommand{\arraystretch}{1.275}
\caption{SIMULATION PARAMETERS}
\vspace{-0.6\baselineskip}
\centering
\begin{tabular}{l l}
\hline
\bfseries Parameter & \bfseries Value\\
\hline\hline
Carrier frequency & 28 GHz\\
System bandwidth & 100 MHz\\
Cell deployment topology & 7-site hexagon\\
Total number of cell $N_\mathrm{cells}$ & 21 \\
Downlink Tx power & 40 dBm\\
Tx (BS) antenna height & 10 m\\
Tx antenna element pattern & Table 7.3-1 in \cite{b10} \\
Tx panel size & 16 $\times$ 8, $\forall b \in \{1,\ldots,8\}$\\
 & 8 $\times$ 4, $\forall b \in \{9,\ldots,12\}$ \\
Tx antenna element spacing & vertical: 0.7$\lambda$\\ 
& horizontal: 0.5$\lambda$\\
Beam elevation angle $\theta_b$ & 90$^{\circ}$, $\forall b \in \{1,\ldots,8\}$ \\
 & 97$^{\circ}$, $\forall b \in \{9,\ldots,12\}$\\
Beam azimuth angle $\phi_b$ & $-$52.5$^{\circ}$$+$15$(b-1)^{\circ}, \forall b \in \{1,\ldots,8\}$\\
& $-$45$^{\circ}$$+$30$(b-9)^{\circ}, \forall b \in \{9,\ldots,12\}$\\
Tx-side beamforming gain model & Fitting model of \cite{b11}\\
Rx (UE) antenna height & 1.5 m\\
Rx antenna element pattern & Isotropic UE: isotropic pattern,\\ 
 & MPUE: based on \cite{b13} \\
Rx panel size  & Single antenna element\\
Rx antenna element antenna gain  & Isotropic UE: 0 dBi\\ 
& MPUE: 5 dBi\\
Total number of UEs $N_\mathrm{UE}$ & 420\\
UE speed  & urban scenario: 60 km/h\\
& highway scenario: 120 km/h\\
Number of simultaneously   & 4\\
scheduled beams per cell $K_b$ & \\ 
CHO preparation offset $o^\mathrm{prep}_{c_0,c^{\prime}}$  & 10\\
CHO execution offset $o^\mathrm{exec}_{c_0,c^{\prime}}$  & 3\\
CHO release offset $o^\mathrm{rel}_{c_0,c^{\prime}}$  & 13\\
CHO replace offset $o^\mathrm{rep}_{c_\textrm{W,S}}$  & 3\\
$T_\mathrm{prep} = T_\mathrm{exec} = T_\mathrm{rem} = T_\mathrm{rep}$ & 80 ms\\
Maximum prepared cells $n_c^{\mathrm{max}}$ & 4\\
Fast-fading channel model & Abstract model of \cite{b11}\\
Time step $\Delta t$, SSB periodicity  & 10 ms, 20 ms\\
Simulated time & 30 s \\
SINR threshold $\gamma_\mathrm{out}$  & $-$8 dB \\
\hline
\end{tabular}
\label{Table1}
\end{table}

A 5G-Advanced network model with an urban-micro (UMi) cellular deployment consisting of a standard hexagonal grid with seven BS sites is considered, each divided into three sectors or cells. The inter-cell distance is 200 meters and the carrier frequency is 28 GHz. 420 UEs are dropped randomly following a 2D uniform distribution over the network at the beginning of the simulation, moving at constant velocities in random waypoint motion along straight lines. Two different mobility scenarios are considered. UEs moving at 60 km/h represent the urban mobility scenario, which is the usual speed limit in the non-residential urban areas of big cities. Whereas UEs moving at 120 km/h represent the highway mobility scenario, which is the usual speed limit on major highways.

\begin{figure}[!b]
\vspace{-1.2\baselineskip}
\centering
\includegraphics[width = 0.96\columnwidth]{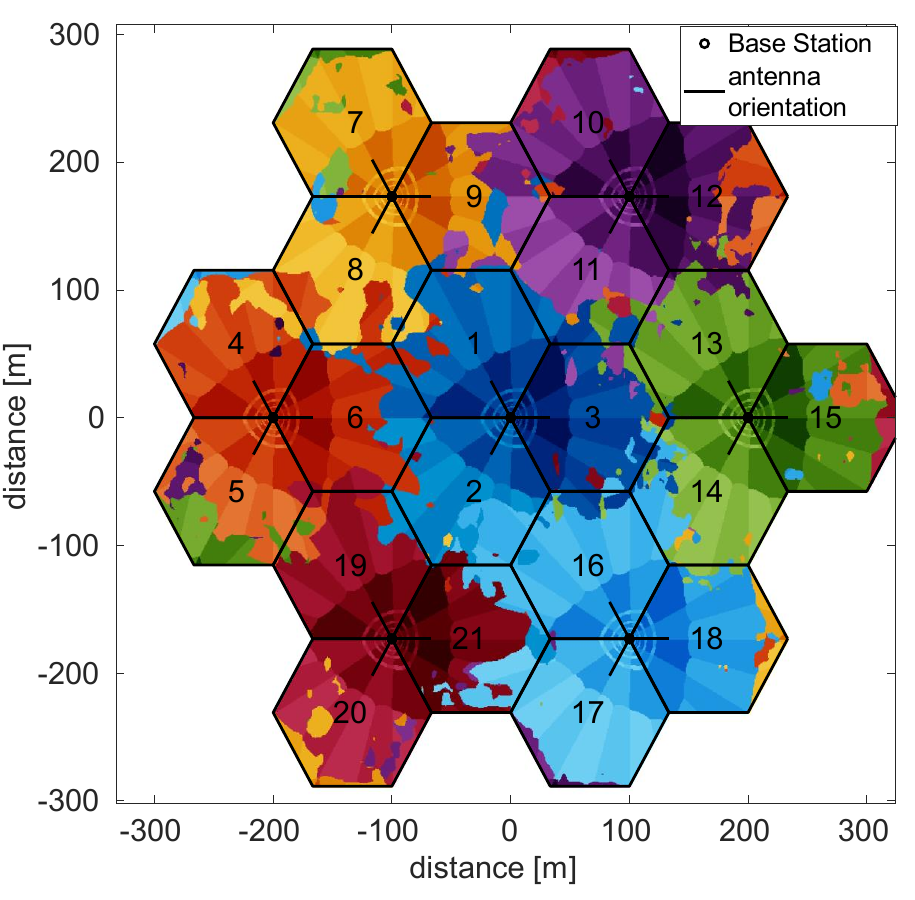}
\vspace{-0.3\baselineskip}
\caption{Simulation scenario consisting of seven hexagonal sites, where each site is serving three cells with 120$^{\circ}$ coverage.} 
\label{Fig5} 
\end{figure}

As per 3GPP's study outlined in Release 15\cite{b10}, the used channel model takes into account shadow fading due to large obstacles and assumes a soft line-of-sight (LOS) for all radio links between the cells and UEs. Fast fading is taken into consideration through the low complexity channel model for multi-beam systems proposed in \cite{b11}, which integrates the spatial and temporal characteristics of 3GPP's geometry-based stochastic channel model (GSCM) \cite{b10} into Jake’s channel model. The transmitter (Tx)-side beamforming gain model is based on \cite{b11}, where a 12-beam grid of beams configuration is considered. $K_b=$ 4 beams are simultaneously scheduled  for all cells in the network. Beams $b \in \{1,\ldots,8\}$ have smaller beamwidth and higher beamforming gain and cover regions further apart from the BS. Beams  $b \in \{9,\ldots,12\}$ have larger beamwidth and relatively smaller beamforming gain and cover regions closer to the BS. This can be seen in Fig. \ref{Fig5}, where the eight outer beams are shown in light color and the four inner beams are shown in dark color. 

Two different UE architectures are considered. The isotropic UE architecture assumes a single antenna element that has an isotropic antenna element radiation pattern, with a gain of 0 dBi \cite{b11}. The MPUE architecture assumes an \textit{edge} design with three directional antenna panels, each with a single antenna element and a maximum gain of 5 dBi \cite{b7},\cite{b8}. The antenna element radiation pattern for each panel is based on \cite{b13}. In line with 3GPP \cite{b9}, MPUE is further divided into two different UE signal measurement schemes. A more detailed explanation of these two different MPUE signal measurement schemes can be found in \cite{b7}. They can be summarized as

\begin{itemize}
\item 
\textit{MPUE-Assumption 3 (MPUE-A3)}: The UE can measure the RSRPs from $c_0$ and neighboring cells on all three panels simultaneously.
\item
\textit{MPUE-Assumption 1 (MPUE-A1)}: The UE can measure the RSRPs from $c_0$ and neighboring cells on just one active panel at a time that follows a round-robin approach.
\end{itemize}

In order to evaluate the instantaneous downlink signal-to-interference-plus-noise ratio (SINR) $\gamma_{c,b}(m)$ of a link between the UE and beam $b$ of cell $c$ the Monte-Carlo approximation given in \cite{b12} for the strict fair resource scheduler is used. This SINR is of key importance in the handover failure model (HOF) model and radio link failure (RLF), both of which are threshold based models dependent on the SINR treshold $\gamma_\mathrm{out}$. The former models failure of a UE to execute a handover from the serving cell $c_o$ to the prepared target cell $c^{\prime}$ and the latter models failure of a UE while still in cell $c_0$. Both these models are explained in detail in \cite{b7}.

\section{Performance Evaluation} \label{Sec5}

In this section the mobility performance of CHO is compared with FCHO for the two UE architectures in question, namely isotropic UE architecture and MPUE architecture. The mobility performance is analyzed in terms of mobility KPIs and signaling overhead for two realistic mobility scenarios.

\subsection{KPIs and CHO Signaling Events} \label{SubSec5.1}
The mobility KPIs used for comparison between FCHO and CHO are explained below.
\begin{itemize}
\item 
\textit{Mobility failures:} Denotes the sum of the total HOFs and RLFs in the network.

\item
\textit{Fast Handover:} Denotes the sum of short-stays and ping-pongs in the network. A short-stay is characterized as a successful handover from one cell to another and then to a third one within a very short time $T_{FH}$ \cite{b14}, e.g., 1 second. Here it is implied that instead a direct handover from the first cell to the third one would have served the purpose. A ping-pong is characterized as a successful handover followed by a handover back to the previous serving cell within $T_{FH}$. Here it is implied that instead potentially both handovers could have been avoided. Although fast handovers are part of successful handovers, they are accounted for as a detrimental mobility KPI that adds additional signaling overhead to the network.

\end{itemize}
Mobility failures and fast handovers are expressed as a percentage of the total number of handover attempts, which is the sum of successful handovers and mobility failures.

\begin{itemize}

\item
\textit{Outage:} Outage is denoted as the time interval during which a UE is not able to communicate with the network and to receive data. This could be due to a number of reasons. Most commonly a UE is assumed to be in outage when the instantaneous SINR $\gamma_{c_0, b_0}$ of the serving cell $c_0$ over the serving beam $b_0$ falls below a SINR threshold $\gamma_\mathrm{out}$. A UE is also assumed to be in outage if either the HOF timer $T_{\mathrm{HOF}}$ or the RLF timer $T_{\mathrm{RLF}}$ expires due to a HOF or RLF, respectively, prompting the UE to initiate a connection reestablishment \cite{b7}. A successful handover, despite being a necessary mobility procedure that allows the UE to remain connected to the network, also contributes to outage since the UE cannot receive data from the network during the time it performs random access to the prepared target cell $c^{\prime}$. When compared to the outage due to connection reestablishment, this outage contribution is modeled as relatively small\cite{b14}. Outage is denoted in terms of a percentage as 
\vspace{-1mm}
\begin{equation}
\label{Eq10} 
\textrm{Outage} \ (\%) = \frac{\sum_{\forall u}{\textrm{Outage duration of UE}} \ u} {N_\mathrm{UE} \ \textrm{x} \ \textrm{Simulated  time}} \ \textrm{x} \ 100. 
\end{equation}
\end{itemize}

The CHO signaling events used for comparison between FCHO and CHO are \textit{CHO Prepare}, \textit{CHO Release} and \textit{CHO Replace}. They reflect the signaling overhead caused by each CHO event and are used to denote the total number of occurrences of each such event discussed in \Cref{Sec2}. All three events are normalized to UE per minute (min).

\subsection{Simulation Results} \label{SubSec5.2}
\begin{figure}[!t]
    \centering
    \subfigure[Mobility KPIs.]
    {
        \includegraphics[width = 0.97\linewidth]{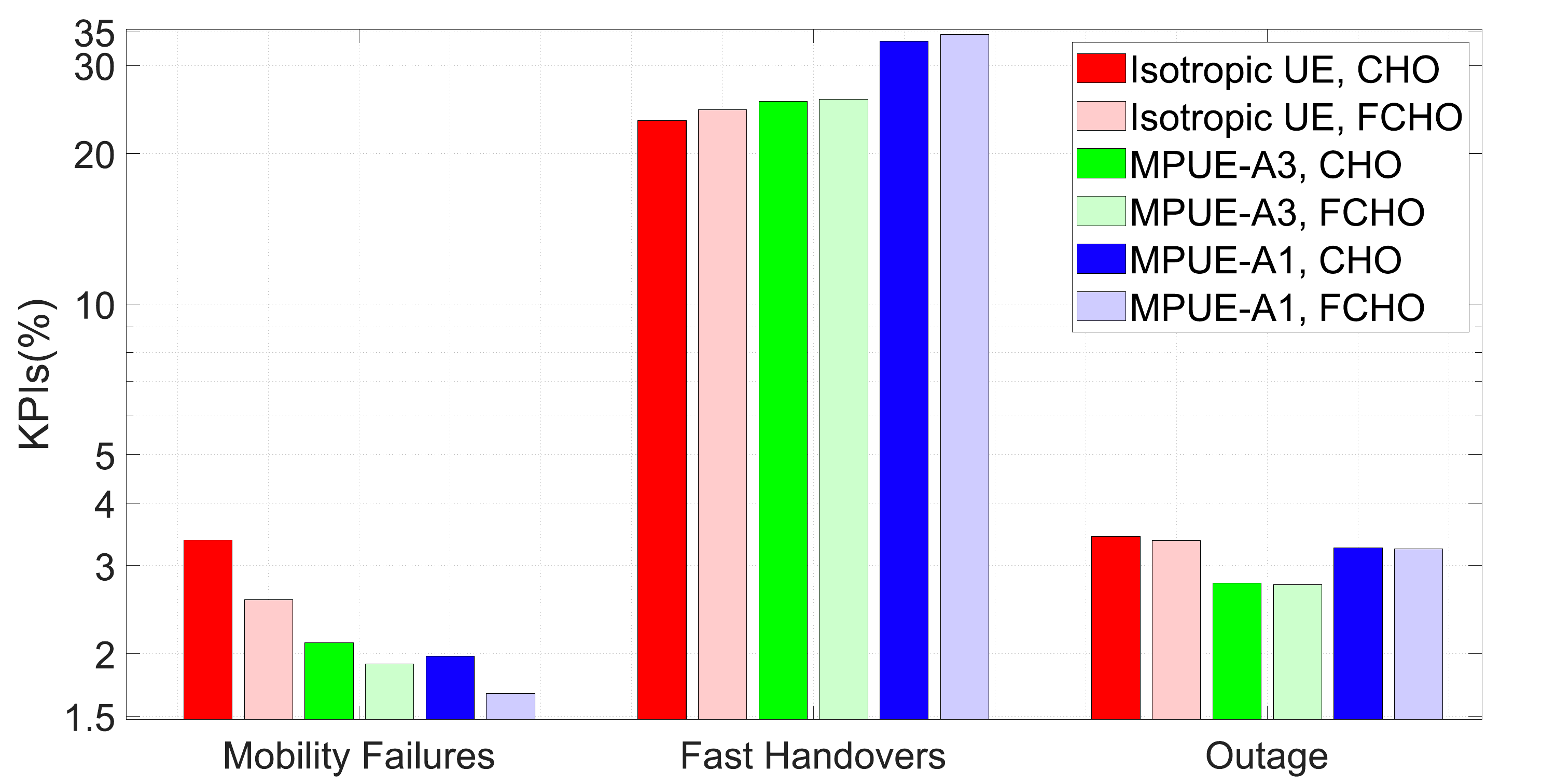}
        \label{fig:Fig6a}
    }
    \\
    \subfigure[CHO signaling overhead.]
    {
        \includegraphics[width = 0.94\linewidth]{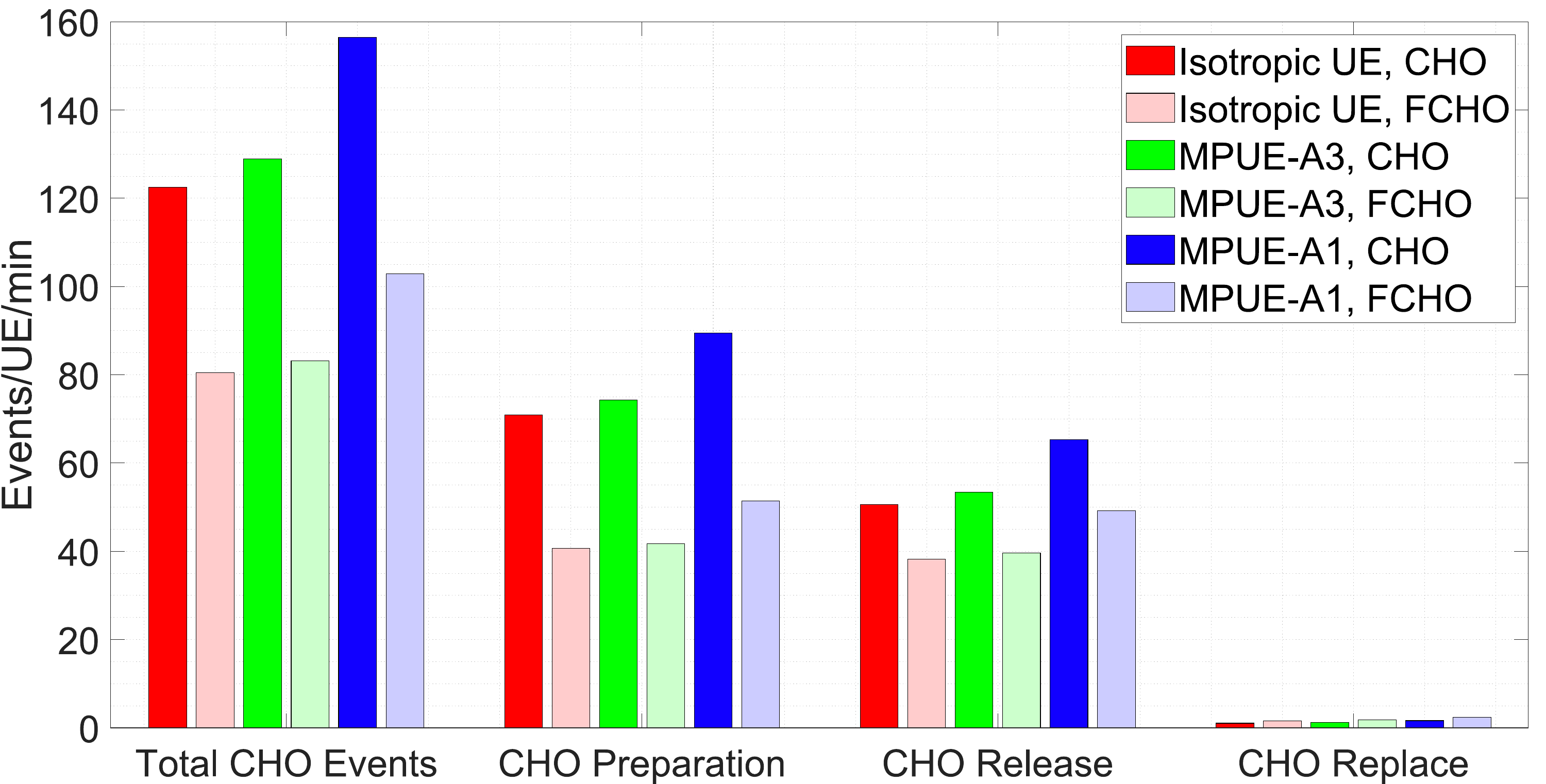}
        \label{fig:Fig6b}
    }
    \vspace{-0.5\baselineskip}
    \caption{For the urban mobility scenario (60 km/h) with $N_\mathrm{UE}$ = 420 (a) Mobility KPIs and (b) CHO signaling overhead.}
    \label{fig:Fig6} 
\end{figure}

Fig. \ref{fig:Fig6} shows the mobility performance for the urban mobility scenario (60 km/h) in terms of mobility KPIs and signaling overhead for CHO (dark color) and FCHO (light color), taken for three different cases, i.e., isotropic UE architecture (in red) and the two MPUE architecture signal measurement schemes, namely MPUE-A3 (in green) and MPUE-A1 (in blue). 

The first key observation from Fig. \ref{fig:Fig6a} in terms of mobility KPIs when FCHO is compared with CHO is that there is a decrease in mobility failures for all three cases considered. For isotropic UE, the mobility failures decrease in absolute percentage terms by 0.8\%. In comparison, for MPUE-A3 and MPUE-A1 the mobility failures decrease by 0.2\% and 0.3\%, respectively. The mobility failures see a greater reduction for isotropic UE since MPUE already addresses some of the corresponding mobility failures on account of a 5 dBi higher directional antenna gain and inter-cell interference suppression from the neighboring cells \cite{b7}\cite{b8}. For isotropic UE, it is seen that fast handovers increase by 1.2\%. In comparison, for MPUE-A3 and MPUE-1 fast handovers increase by 0.3\% and 1\%, respectively. It can be inferred that some of the mobility failures in CHO have turned into fast handovers in FCHO. For instance, for MPUE-A3 (in green) it can be said that the 0.2\% decrease in mobility failures correspondingly increases fast handovers by 0.2\%, with the remaining 0.1\% coming from normal successful handovers that have resulted in fast handovers on account of immediate handover executions that has resulted in unreliable handover decisions. It is also worthwhile to note that the problem of high number of fast handovers in MPUE-A1 becomes more pronounced with FCHO because of the use of the round-robin approach  for panel switching that results in outdated measurements on two out of the three panels and triggers some unnecessary handovers \cite{b7}. 

When FCHO is compared with CHO for the three cases it is seen that the outage is nearly equal ($<$ 0.1\%). This is because the decrease in the outage contribution due to the reduction in mobility failures is offset by the outage contribution increase due to relatively higher increase in fast handovers. It can be concluded that FCHO is advantageous over CHO because minimizing mobility failures  has a higher priority  than minimizing fast handovers \cite{b14}. 

For signaling overhead shown in Fig. \ref{fig:Fig6b} it is seen that there is a significant reduction in the total number of CHO events for all three cases. It is also seen that the total CHO events are comparable to fast handovers in \ref{fig:Fig6a}. In relative percentage terms, for isotropic UE the total CHO events decrease by 34.3\% when FCHO is compared with CHO. For MPUE-A3 and MPUE-A1 a comparable reduction of 35.5\% and 34.2\% is seen. This reduction stems from the reduction in both CHO preparation and CHO release events. It is also observed that there is a higher reduction in CHO preparation events as compared to CHO release events for each case. For instance, for MPUE-A3 (in green) the CHO preparation events reduce  by 43.7\%, whereas the CHO release events reduce by 25.7\%. As discussed in \Cref{Sec3}, this is because in FCHO some of the prepared target cells that are retained after a successful handover become outdated and need to be released from the prepared cell list over time. But this is offset by the relatively larger reduction in CHO release events, which is a consequence of the reduction in CHO preparation events in FCHO. Lastly, it can be seen that there is a relatively small increase in CHO replace events for all the three cases considered. This is because in FCHO the RSRPs of the prepared target cells weaken over time in more instances due to some prepared cells becoming outdated. 

\begin{figure}[!b]
    \centering
    \subfigure[Mobility KPIs.]
    {
        \includegraphics[width = 0.96\linewidth]{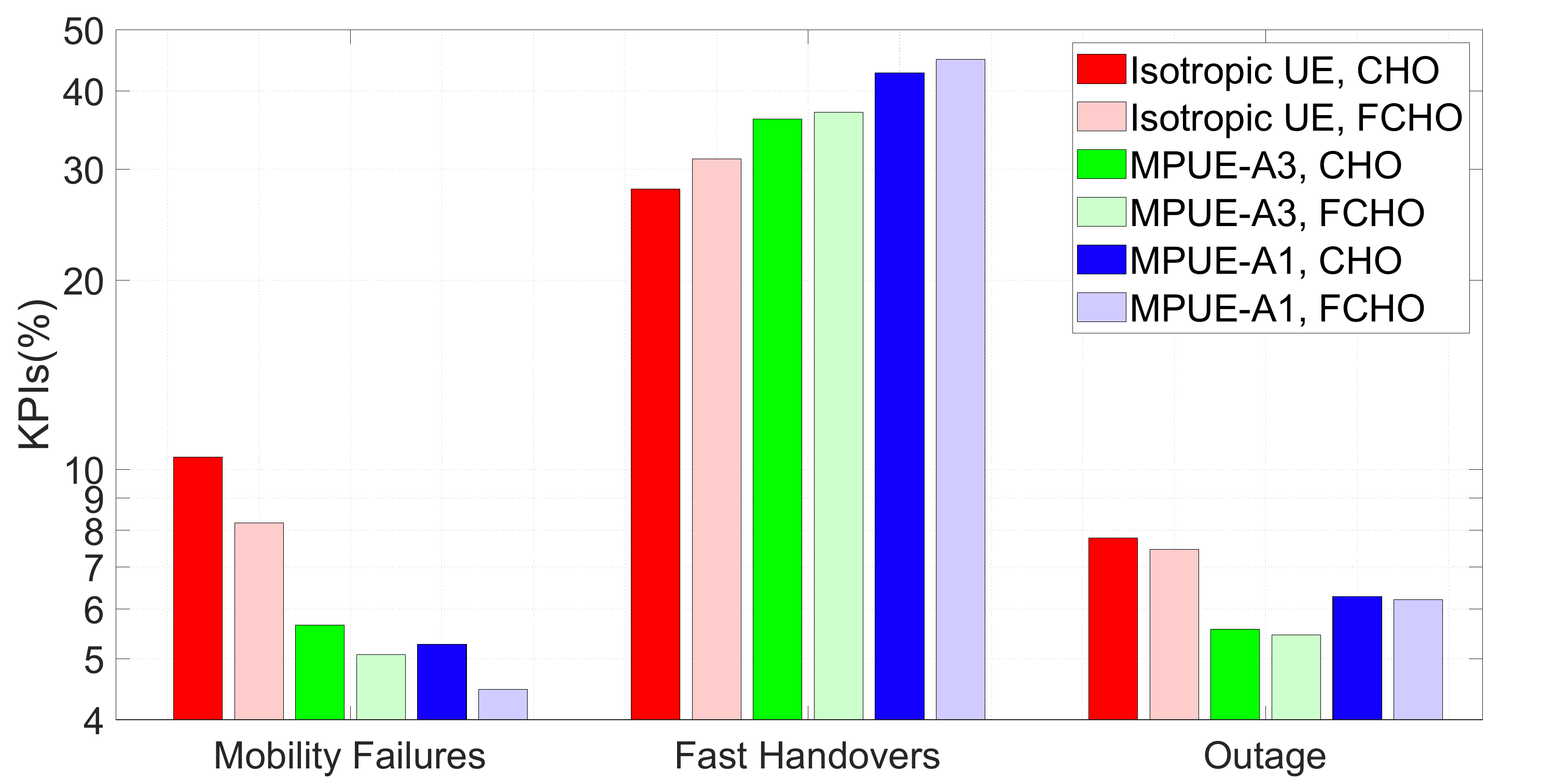}
        \label{fig:Fig7a}
    }
    \\
    \subfigure[CHO signaling overhead.]
    {
        \includegraphics[width = 0.94\linewidth]{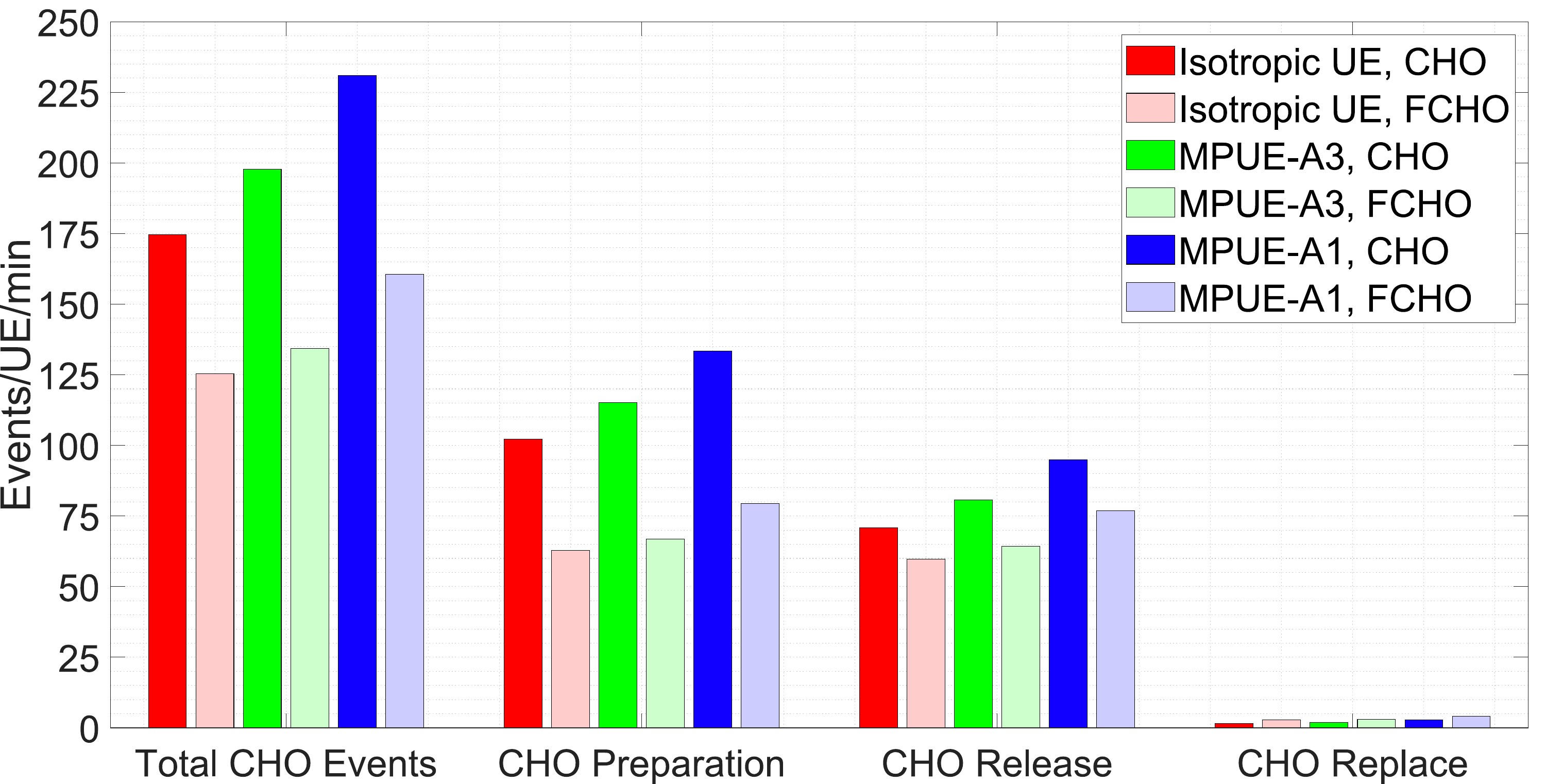}
        \label{fig:Fig7b}
    }
    \vspace{-0.2\baselineskip}
    \caption{For the highway mobility scenario (120 km/h) with $N_\mathrm{UE}$ = 420 (a) Mobility KPIs and (b) CHO signaling overhead.}
    \label{fig:Fig7}
\end{figure}

The mobility KPIs are shown in Fig. \ref{fig:Fig7a} for the highway mobility scenario. At 120 km/h mobility becomes more challenging because of greater temporal variations in the signal RSRPs due to dominant fast fading. Additionally, the UEs traverse more cell boundaries compared to 60 km/h and therefore the probability of mobility failures also increases. However, this also means that FCHO can be more beneficial at higher UE speeds in terms of addressing these mobility problems. For example, for isotropic UE, the mobility failures reduce by 2.3\%. For MPUE-A3 and MPUE-A1 the reduction is by 0.6\% and 0.9\%. When compared to \ref{fig:Fig6a} the decrease is almost 3x for all the three cases. It is also seen that as consequence of traversing more cell boundaries, more fast handovers occur than compared to the corresponding cases in Fig. \ref{fig:Fig6a}. This also translates into greater percentage increase in fast handovers with FCHO. When FCHO is compared with CHO, it is seen that the outage is comparable ($<$0.3\%) because the outage contribution decrease due to less mobility failures is offset by the outage contribution increase due to increase in fast handovers. For the signaling overhead in Fig. \ref{fig:Fig7b} it can be observed that the total CHO events are significantly higher than for the urban mobility scenario in Fig. \ref{fig:Fig6b} due to higher fast handovers. The gain in signaling overhead by the reduction in total CHO events is comparable to \ref{fig:Fig6b}, e.g. 32.1\% relative reduction in for MPUE-A3 (in green) in \ref{fig:Fig7b} as compared to 35.5\% in \ref{fig:Fig6b}.

\section{Conclusion} \label{Sec6}
In this paper the performance of FCHO for MPUE and isotropic UE architectures has been analyzed in a multi-beam network. FCHO is an essential part of 5G-Advanced and this paper is a novel attempt to understand the mobility performance and associated challenges of FCHO with advanced UE architectures. Two different mobility scenarios have been considered and it is seen that both MPUE schemes perform considerably better compared to their CHO counterparts. In the urban mobility scenario, FCHO with MPUE-A3 reduces mobility failures by 0.2\%, while significantly reducing the signaling overhead by 35.5\%. In the highway mobility scenario the corresponding gains are 0.6\% and 32.1\%, respectively. This is because in FCHO the handovers can be executed immediately without any additional repreparation signaling due to reutilization of target cell preparations. Based on these findings, further studies may be carried out to optimize the resource reservation time as one downside of FCHO is that resources are reserved for longer time durations \cite{b3}.


\end{document}